\documentclass[12pt]{iopart}

\usepackage{a4}

\usepackage{iopams}  
\usepackage{bm}
\usepackage{epsfig,color}
\usepackage{amstext}
\usepackage{amssymb}
\usepackage[latin1]{inputenc}
\usepackage{mathptmx}

\DeclareSymbolFont{letters}{OML}{txmi}{m}{it}

\newcommand{\eqref}[1]{(\ref{#1})}

\begin{document}

\title{Absorbing boundary conditions for dynamical many-body quantum systems}

\author{S{\o}lve Selst{\o}}
\address{Centre of Mathematics for Applications, University of Oslo, N-0316 Oslo, Norway}
\ead{solve@cma.uio.no}
\author{Simen Kvaal}
\address{Centre of Mathematics for Applications, University of Oslo, N-0316 Oslo, Norway}
\ead{simenkva@matnat.uio.no}

\newcommand{\ket}[1]{|#1\rangle}
\newcommand{\bra}[1]{\langle#1|}
\newcommand{\braket}[1]{\langle#1\rangle}

\begin{abstract}
In numerical studies of the dynamics of unbound quantum mechanical systems, absorbing boundary conditions are frequently applied. Although this certainly provides a useful tool in facilitating the description of the system, its applications to systems consisting of more than one particle is problematic. This is due to the fact that all information about the system is lost upon absorption of one particle; a formalism based solely on the Scrh{\"o}dinger equation is not able to describe the remainder of the system as particles are lost. 
Here we demonstrate how the dynamics of a quantum
system with a given number of identical fermions may be described in a manner
which allows for particle loss. A consistent formalism which
incorporates the evolution of sub-systems with a reduced number of
particles is constructed through the Lindblad equation.
Specifically, the transition from an $N$-particle system to an 
$(N-1)$-particle system due to a complex absorbing 
potential is achieved by relating the Lindblad operators to annihilation 
operators.
The method allows for a straight forward interpretation of how many constituent particles have left the system after interaction.
We illustrate the formalism using one-dimensional two-particle model
problems.
\end{abstract}




\maketitle

\section{Introduction and basic theory}

Large efforts have been invested in understanding the quantum
mechanical behaviour of dynamical unbound systems involving several
particles. Experimental advances allow us to study processes in which
internal interactions between the constituent particles play a crucial
role -- in addition to dynamics induced by external
perturbations. Examples of such processes 
may be 
photo ionization of helium
\cite{Taylor1996} and cascaded Auger processes following ionization of
inner core electrons in atoms \cite{Krausz2007}.
Of course, in order to understand such
processes, theoretical and numerical studies are
required. Furthermore, the theoretical interest of such systems, be it
within the context of solid state, molecular, atomic or nuclear
physics, is spurred by the fact that they represent demanding tasks.
One obvious challenge in describing unbound systems is that their spatial extension may become arbitrarily large. 
Moreover, even if one is able to represent
the whole system within a finite space, extracting relevant
information from the final wave function may be far from trivial.

The unbound quantum systems under study may often be thought of as having an interaction region of finite spatial extension and an asymptotic region where the unbound part of the system travels outwards. As this asymptotic behaviour often is well known, it may be desirable to describe only the dynamics of the part of the wave function belonging to the interaction region. In a numerical implementation this cannot be done by simply resorting to a representation of space that is smaller than the extension of the wave function as this leads to unphysical reflections at the boundary. However, it can be achieved by imposing {\it absorbing boundary conditions}, i.e.~by demanding that the wave function vanishes as the particle approaches the edge of the numerical grid -- 
preferably with as little reflection as possible, see references~\cite{Kosloff1986,Fevens1999} or the recent review of reference~\cite{Antoine2008}. Such absorbers are frequently referred to as {\it perfectly matched layers} in the context of general wave equations \cite{Berenger1994}. 

When propagating a wave packet on a numerical grid, a common way to 
impose absorbing boundary conditions is by adding a complex absorbing potential (CAP) to the Hamiltonian of the system \cite{Manolopoulos2002}. Complex absorbing potentials are widely used e.g. within molecular dynamics \cite{Monnerville1999,Grozdanov2006,Karlsson2009} and atomic physics \cite{Kulander1987,Protopapas1996}. Alternatively, this way of absorbing particles may be formulated as multiplying the wave function with a masking function at each time step \cite{Taylor1996}.

In any case the resulting effective Hamiltonian acquires an anti-Hermitian part,
\begin{equation}
\label{eq:Heffective}
H_{\rm eff} \equiv H - i \Gamma,
\end{equation}
where both $H$ and $\Gamma$ are Hermitian and $\Gamma$ is positive
semi-definite ($\Gamma\geq 0$).

The wave function $\ket{\Psi(t)}$ of the
system obeys the Schr{\"o}dinger equation
\begin{equation}
  i \hbar \frac{\rmd}{\rmd t} \ket{\Psi(t)} =
    H_\text{eff}\ket{\Psi(t)}.
  \label{eq:schroedinger}
\end{equation}
A density operator $\rho(t)$
correspondingly obeys the von Neumann equation
\begin{equation}
  i \hbar \frac{\rmd}{\rmd t} \rho(t) = H_\text{eff}\rho(t) -
  \rho(t)H_\text{eff}^\dag = [H,\rho(t)] - i \{\Gamma,\rho(t)\}.
  \label{eq:vonneumann}
\end{equation}
It is easy to see that the evolution is non-unitary, viz,
\begin{equation}
  \frac{\rmd}{\rmd t}\Tr \left[ \rho(t) \right] = 
-\frac{2}{\hbar} \Tr \left[ \Gamma\rho(t) \right] \leq 0,
\end{equation}
so that probability is ``lost'' if the density matrix overlaps with
anti-Hermitian part of the effective Hamiltonian.

Although methods involving non-Hermitian effective Hamiltonians may come in very handy in reducing the complexity in
describing potentially unbound systems, there is one major problem when the system contains more than one particle: As one
particle leaves the rest of the system and is subsequently absorbed, the
entire wave function is lost. This is obvious since the wave function
is normalized to the probability of finding \emph{all} particles
within the space defined by the numerical implementation. 
In other words: If an initial $N$-particle system
gradually ``loses'' one particle through some non-Hermitian
``interaction'', the corresponding wave function $\ket{\Psi_N(t)}$ gradually goes
to zero -- {\it not} to some wave-function corresponding to $N-1$
particles.

As it is desirable to be able to describe dynamics where several
particles are lost one by one, one may try and construct a formalism
where an $(N-1)$-particle wave function $\ket{\Psi_{N-1}(t)}$ is created
as one particle is lost, and where this wave function may be propagated
using the corresponding Schr{\"o}dinger equation including some source
term. Once being able to do this, the extension to $N-2$ particles
etc.~follows by induction. However, as it turns out, such a
construction is not possible due to the fact that the process of
losing particles in this way is irreversible: As a particle is absorbed,
information is irretrievably lost. 
Hence, a Markovian master equation should be a more suitable starting
point than a pure state approach. Lindblad \cite{Lindblad1976} was
able to show that in order for the evolution of a system to be
Markovian, trace-conserving and positive, the density operator
$\rho(t)$ has to obey an equation of the form
\begin{equation}
\label{Lindblad}
i\hbar \frac{\rmd}{\rmd t} \rho = \left[H, \rho \right] -
\mathcal{D}(\rho) 
\end{equation}
with the so-called Lindbladian $\mathcal{D}(\rho)$ given by the
generic expression \cite{Schlosshauer}
\begin{equation}
\label{eq:lindbladian}
\mathcal{D}(\rho) = i \sum_{m,n} \gamma_{m,n} \left(A_m^\dagger A_n \rho + \rho A_m^\dagger A_n  - 2 A_n \rho A_m^\dagger \right).
\end{equation} 
Here, $H$ is the Hamiltonian governing the unitary part of the
evolution. The operators $A_n$ are referred to as Lindblad operators.
In a diagonal representation, the Lindbladian simplifies to
\begin{equation}
\label{eq:lindbladian_diag}
\mathcal{D}(\rho) = i \sum_n \left(A_n^\dagger A_n \rho + \rho A_n^\dagger A_n  - 2 A_n \rho A_n^\dagger \right).
\end{equation} 
Equation \eqref{Lindblad} is usually used to describe energy
dissipation to the environment \cite{Sandulescu1987}. However, it has
recently been demonstrated that the spontaneous decay of unstable particles may also be described within this formalism \cite{Caban2005,Bertlmann2006}. In reference~\cite{Caban2005} a master equation of Lindblad from is obtained for a Hilbert space consisting of unstable particle states and vacuum, however without any dynamical degrees of freedom. The decay rates of the unstable particles are introduced as parameters. Similarly, in reference~\cite{Bertlmann2006} decay from one system to another one consisting of its decay products is described by introducing a single Lindblad operator.
In these works it is demonstrated that one may describe decoherence
along with decay in this context.

In the present work these ideas have been used to generalize the
standard one-particle absorbing boundary technique to an $N$-body
setting, and it is demonstrated that this is indeed the \emph{natural}
way to do this. Alternatively, in a more general context, this
formalism allows the study of a system which gradually loses particles
to some environment due to the non-Hermitian part of $H_\text{eff}$.

We comment that the absorption process does not affect the dynamics on its
interior. However, removing a particle may still affect the other
particles via their interaction.  
Hence, in order to obtain physically meaningful results, any
dependence of the absorber must be eliminated by placing the absorber
sufficiently far away from the interaction region -- as is the case
for any implementation featuring absorbing boundary conditions.  

In the following section, section~\ref{sec:FockSpace}, the formalism will be formulated for identical fermions exposed to a complex absorbing potential. 
In section~\ref{sec:example} two examples featuring two identical fermions in one dimension are given. Finally, in section~\ref{sec:conclusion}, conclusions are drawn and a few future perspectives are outlined.

\section{Fock space description}
\label{sec:FockSpace}

The natural setting for describing a system of a variable number of
fermions is Fock space, the direct sum of all $n$-fermion spaces
$\mathcal{H}_n$. As we wish to describe at most $N$ particles, it
suffices to consider
\begin{equation}
  \mathcal{H} = \bigoplus_{n=0}^N \mathcal{H}_n.
\end{equation}
An arbitrary $n$-fermion state $\ket{\Psi}\in\mathcal{H}_n$ can be
written
\begin{equation}
  \ket{\Psi} = \frac{1}{n!}\int \rmd x^n \, \Psi(x_1\cdots x_n)
  \boldsymbol{\psi}^\dag(x_1)\cdots\boldsymbol{\psi}^\dag(x_n) \ket{-},
\end{equation}
where the field operator $\boldsymbol{\psi}^\dag(x)$ creates a particle at position
$x$. 
The field operators obey the usual anti-commutator relations
\begin{equation}
  \{ \boldsymbol{\psi}(x), 
\boldsymbol{\psi}(x') \} 
= 0, \qquad 
  \{ \boldsymbol{\psi}(x), \boldsymbol{\psi}^\dag(x')\} = \delta(x-x').
\end{equation}
Here, $x$ may denote \emph{all} degrees of freedom associated with a
particle, including eigenspin. Moreover, $\Psi(x_1\cdots x_n)$ is the
anti-symmetric local wave function of the
$n$-particle system, and $\ket{-}\in\mathcal{H}_0$ is the vacuum
state, containing zero particles.

A system of $n$ identical fermions interacting with at most two-body
forces has the Hamiltonian
\begin{equation}
\label{eq:GeneralHamiltonian}
  H = \sum_{i=1}^n h(x_i) + \sum_{j<i}^n u(x_i,x_j).
\end{equation}
Here, $h(x_i)$ (resp. $u(x_i,x_j)$) is a one-body (two-body) operator acting on the degrees-of-freedom associated with particle $i$ (and $j$). Using
field operators, the Hamiltonian can be written
\begin{eqnarray}
\label{Hamiltonian2ndQuant}
  H &=& \int \rmd x \, \boldsymbol{\psi}^\dag(x) h(x) \boldsymbol{\psi}(x) +
 \\
\nonumber
  & & \frac{1}{2}\int \rmd x \rmd x'\,
 \boldsymbol{\psi}^\dag(x)\boldsymbol{\psi}^\dag(x') u(x,x')
  \boldsymbol{\psi}(x')\boldsymbol{\psi}(x),
\end{eqnarray}
the point here being that $H$ given on this form is \emph{independent}
of the number of particles in the system. The exact expression is
irrelevant at this point.
The CAP, which is diagonal in $x$, 
is conveniently introduced as
\begin{equation}
\label{Absorber2ndQuant}
  -i\Gamma = -i \int \rmd x \, \boldsymbol{\psi}^\dag(x) \Gamma(x)
  \boldsymbol{\psi}(x).
\end{equation}

By comparing (\ref{Lindblad}), (\ref{eq:lindbladian_diag}) and (\ref{Absorber2ndQuant}) with the von Neumann equation~(\ref{eq:vonneumann}), we see that the Lindblad operators, for a diagonal representation of the Lindbladian, must fulfil
\begin{equation}
\label{LindbladAbs}
\Gamma = \int \rmd x \; A^\dag(x) A(x)
\end{equation}
in order to reproduce the anti Hermitian ``interaction'' leading to absorption. Here we have allowed for an integral instead of a sum in (\ref{eq:lindbladian_diag}).
Furthermore, the last term of the Lindbladian, which is absent in the von Neumann equation, $-2 A_n \, \rho A_n^\dagger \rightarrow -2 A(x) \rho A^\dagger(x)$, should map an $N$-particle system into an $(N-1)$-particle system. Hence, $A(x)$ should map
$\mathcal{H}_N$ into $\mathcal{H}_{N-1}$, which means that the Lindblad operator $A(x)$ must be on the form
\begin{equation}
\label{LindbladNonDiag}
A(x) = \int \rmd x \, a(x,x') \boldsymbol{\psi}(x'). 
\end{equation}
The simplest choice that satisfies (\ref{LindbladAbs}) is the diagonal form
\begin{equation}
\label{LindbladDef}
A(x) \equiv \sqrt{\Gamma(x)} \boldsymbol{\psi}(x).
\end{equation}
We will return to the justification of why this is the proper way to define the Lindblad operators shortly.

With (\ref{LindbladDef}), the Lindbladian~(\ref{eq:lindbladian_diag}), may
immediately be written
\begin{eqnarray}
  \mathcal{D}(\rho) = i\Gamma\rho + i\rho\Gamma - 2 i\int dx\; \Gamma(x)
 \boldsymbol{\psi}(x)\rho \boldsymbol{\psi}^\dag(x),
\end{eqnarray}
and the master equation \eqref{Lindblad} becomes
\begin{equation}
  i\hbar \frac{\rmd}{\rmd t}\rho = [H,\rho] - i\{\Gamma,\rho\}
 +2i\int \rmd x\;
  \Gamma(x) \boldsymbol{\psi}(x)\rho\boldsymbol{\psi}^\dag(x).
  \label{eq:main-master}
\end{equation}
This is our fundamental dynamical formulation of particle loss due to a CAP. 

\begin{figure}
\begin{center}
\epsfysize=5.5cm \epsfbox{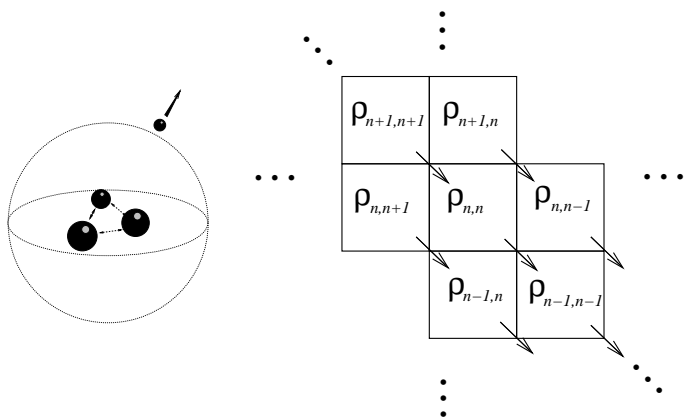}
\end{center}
\caption{Illustration of how the various blocks of the density matrix are populated via the one above to the left only. As a consequence of this, when the density matrix is block diagonal, this diagonal structure is maintained, and losing a particle (left figure) corresponds to moving downwards along the block-diagonal of $\rho$.}
\label{fig:blocks}
\end{figure}

Let us consider the master equation \eqref{eq:main-master} in some
detail. We may partition the density matrix $\rho$ into blocks, viz,
\begin{equation}
  \rho = \sum_{n=0}^N \sum_{m=0}^N \rho_{n,m},
\end{equation}
where $\rho_{n,m} = P_n \, \rho P_m$, with $P_n$ being the orthogonal
projector onto $\mathcal{H}_n$.  Each block evolves according to
\begin{eqnarray}
  \label{eq:block-flow}
  i\hbar \frac{\rmd}{\rmd t}\rho_{m,n} &=& [H,\rho_{m,n}] -
  i\{\Gamma,\rho_{m,n}\} + \\ & & 2i\int \rmd x\; 
  \Gamma(x) \boldsymbol{\psi}(x)\rho_{m+1,n+1}\boldsymbol{\psi}^\dag(x),
\nonumber
\end{eqnarray}
showing that the flow of $\rho_{n,m}$ depends on
that of $\rho_{n+1,m+1}$, but not the other way around. This is
illustrated in figure~\ref{fig:blocks}. 
Also, notice that $\rho_{N,N}$ obeys
the original von Neumann equation \eqref{eq:vonneumann} as there are
no couplings to other blocks in this case as particles do not enter
the $N$-particle system.

We now return to the question of whether the definition~(\ref{LindbladDef}) of the Lindblad operators, which obviously is the simplest one, is the only adequate one.
Indeed, during the transition from an $N$-particle system to an $(N-1)$-particle system, only the unabsorbed part of the original system should be reproduced within the $(N-1)$-particle system, and this leads to the above choice. 
To see this, consider a somewhat idealized example consisting of two non-interacting particles. Since they do not interact, their wave function is given by a Slater determinant at all times, 
$\Psi = [\alpha(x_1,t) \beta(x_2,t) - \beta(x_1,t) \alpha(x_2,t)]/\sqrt{2}$.
The state $\alpha$ does not overlap with the absorber at any time, i.e.
\begin{equation}
\label{NoOverlap}
\Gamma(x) \alpha(x,t) = 0, \quad\text{for all $x$ and $t$},
\end{equation}
and $\beta$ corresponds to an unbound particle travelling outwards.
We suppose that as $t$ approaches infinity, the particle in the state $\beta$ is completely absorbed, and our numerical representation of the final system should converge towards the pure one-particle state $\alpha$. 

Indeed, the evolution dictated by (\ref{eq:block-flow}) follows this pattern. 
The block of the density matrix corresponding to the two-particle system, $\rho_{2,2}$, remains a pure state -- albeit with decreasing norm as $\beta$ is absorbed --  and 
the evolution of the one-particle block $\rho_{1,1}$ simplifies due to (\ref{NoOverlap}) to
\begin{equation}
\label{TrivialOnePartEvolution}
i \hbar \dot{\rho}_{1,1} = [h, \rho_{1,1}] + 2 i \int \Gamma(x) |\beta(x; t)|^2 \, \rmd x \; |\alpha \rangle \langle \alpha |.
\end{equation} 
Hence, the one-particle part of the system is always proportional to a pure $\alpha$-state, and, since the trace of the entire system is conserved, this simply integrates to $\rho_{1,1}(t \rightarrow \infty) = |\alpha \rangle \langle \alpha|$, as it should. 

On the other hand, with a non-diagonal form of the Lindblad operators, c.f. (\ref{LindbladNonDiag}), we would have contributions to $\dot{\rho}_{1,1}$ of the form $|\beta \rangle \langle \alpha|$ and its Hermitian adjoint in addition to the pure state contribution. These contributions clearly cannot be allowed, as $\rho_{1,1}$ should be independent of the state $\beta$ of the outgoing particle. Hence, we find that, up to an arbitrary phase factor, the definition~(\ref{LindbladDef}) is the proper way to define $A(x)$.

For a non-interacting $N$-fermion system where $N_\alpha$ particles are bound and $N_\beta$ particles are unbound, and with an initial state given by the Slater determinant 
$|\Psi\rangle = | \{ \alpha_1,...,\alpha_{N_\alpha},\beta_1,...,\beta_{N_\beta} \} \rangle$, it may be verified by inspection that the source term for the $N_\alpha$-th block, $\rho_{N_\alpha,N_\alpha}$, is always proportional to 
$|\{ \alpha_1,...,\alpha_{N_\alpha} \} \rangle 
\langle \{ \alpha_1,...,\alpha_{N_\alpha} \} |$ given that none of the
$\alpha_i$-states overlap with $\Gamma$. The intermediate density
matrix blocks $\rho_{N - n,N - n}$, where $0 < n < N_\beta$, will
approach zero as $t\rightarrow\infty$, but transiently describe (mixed)
states where $n$ particles have left. 

Of course, in the more interesting context of interacting particles, the structure of the diagonal blocks of the density operator, $\rho_{n,n}$, is more complex than in the special case of non-interacting particles.

\subsection{Consequences and interpretation}

Typically, our starting point is a pure $N$-particle state, $\ket{\Psi(0)}\bra{\Psi(0)}$, in which case \eqref{eq:block-flow} reduces to the ordinary Schr{\"o}dinger equation \eqref{eq:schroedinger}, 
which was our original formulation. Moreover, it is easy to see that
$\rho(t)$ will remain block diagonal for all $t$ in this case, i.e.~$\rho_{n,m}=0$ if
$n\neq m$. But \eqref{eq:block-flow} shows that $\rho_{n, n}(t)$ in
general is a \emph{mixed state}, unlike
$\rho_{N, N}(t)=\ket{\Psi(t)}\bra{\Psi(t)}$. 
This is due to the information loss when admitting ignorance of the
whereabouts of the removed particle.

We stress that by construction, $\Tr [\rho(t)] = 1$ for all
$t$. Probability flows monotonically from $\mathcal{H}_N$ into
$\mathcal{H}_{n}$, and the particle absorbed from $\rho_{N,N}$ is not
present in $\rho_{n,n}$, but is erased, and $\rho_{n, n}$ is a proper
description of an $n$-fermion system. In this way, we see that the
above construction is a \emph{natural} generalization of the original
non-Hermitian dynamics, which describes the classical removal of a
particle, into one that also describes the remaining system in a
consistent way. It is striking to notice that the CAP, $-i\Gamma(x)$, is
\emph{already given} as one of the terms in the Lindbladian, so that
the Fock space formulation follows almost immediately.

It is worthwhile to mention that the traces of the blocks $\rho_{n,n}$ along the diagonal of $\rho$ have very simple interpretations as the probability $P(n)$ of having $n$ particles in the system upon a measurement, i.e.
\begin{equation}
\label{PartialTrace}
  P(n) \equiv \Tr_n [\rho(t)] \equiv \Tr [\rho_{n,n}(t)].
\end{equation}
In particular, $P(N) = \braket{\Psi|\Psi} \le 1$, and
$P(0) = 1-\sum_{n=1}^N P(n)$. 
Hence, within this formalism, distinguishing between single, double etc.~ionization of atoms and molecules is straightforward.

For any observable $A$ the expectation value is given by
$\braket{A}\equiv\Tr[A\rho]=\sum_n\Tr[A \rho_n]$. 
For example, the expected number of particles in the system is given by
\begin{equation}
 \braket{\mathcal{N}} = \int \rmd x\,
 \Tr[\boldsymbol{\psi}^\dag(x)\boldsymbol{\psi}(x)\rho ] =
 \sum_{n=1}^N n P(n).
\end{equation}
It may also be useful to consider conditional expectation values:
\begin{equation}
\label{Conditional}
  \braket{A}_{n} \equiv
  \frac{\Tr_n(A\rho)}{\Tr_n(\rho)},
\end{equation}
i.e., the expectation value of $A$ \emph{given} that the system is
found in an $n$-particle state.

Obviously, as particles are ``removed'' by the absorber, information
is lost. This information loss may be quantified by the von Neumann
entropy, $S \equiv -\Tr[\rho \log \rho]$,  
or the closely related notion of \emph{purity}, defined by \cite{Schlosshauer} 
\begin{equation}
\label{Purity}
\varsigma \equiv \Tr(\rho^2) \le 1.
\end{equation}
Unity minus this quantity, $1-\varsigma$, is a measure of the amount of mixedness, and the purity $\varsigma$ is 1 for pure states only.
Similar to conditional expectation values, c.f. (\ref{Conditional}), one may define conditional purity and von Neumann entropy as the corresponding quantity of the re-normalized block, i.e. $\rho_{n,n}/\Tr[\rho_{n,n}]$.

Of course, for a full quantum mechanical description in terms of the
complete (unabsorbed) $N$-particle wave function, no information is
lost. 
Indeed, the absorption of
particles is a \emph{semi-classical concept}. In the full
$N$-fermion quantum system no such separation is possible. On the other hand, as
the Schrödinger equation for the $N$ particles is separable in the
non-interacting $\alpha_i$ and $\beta_i$ states discussed above,
giving a Slater determinant of the time evolved one-particle states as
the full solution, the Lindblad equation 
is seen to correctly 
construct the $N_\alpha$-particle Slater determinant resulting from
the removal of the $N_\beta$ outgoing particles from this Slater
determinant. Thus, the Lindblad equation exactly encapsulates the
approximate separation of non-interacting
quantum systems far apart.

\section{Example: Two identical spin $\frac{1}{2}$-particles in one dimension}
\label{sec:example}

In the following we consider two fermions in one dimension with
the one-body Hamiltonian $h$ given by
\begin{equation}
\label{OnePartHam}
  h = -\frac{\hbar^2}{2m} \frac{\partial^2}{\partial x^2} + V(x,t),
\end{equation}
where $V(x,t)$ is some external, possibly time-dependent, potential. The fermions interact via a
potential $U(x_1-x_2)$ and the extension of the system is effectively reduced by imposing a CAP. 
With two interacting identical fermions, we are confined to the
Hilbert space
\begin{equation}
\label{Hilbert}
{\cal H}={\cal H}_2 \oplus {\cal H}_1 \oplus {\cal H}_0.
\end{equation}

We discretize this system using a uniform grid in the interval 
$[0,x_{\rm max})$ containing $N$ points $\{x_j\}_{j=0}^{N-1}$, $x_j= j h$, with
$h = x_{\rm max}/N$.
The field operators $\boldsymbol{\psi}^\dag(x)$ can be
replaced by a finite number of creation operators $c_j^\dag$,
creating a particle at grid point $x_j$. 
These operators, which obey the
usual fermion anti-commutator rules
\begin{equation}
  \{ c_j, c_k \} = 0, \qquad \{c_j,c^\dag_k\} = \delta_{j,k},
\end{equation}
map discrete anti-symmetrized $\delta$-function bases for $\mathcal{H}_n$ into bases for
$\mathcal{H}_{n\pm 1}$.

The Hamiltonian takes the form
\begin{equation}
  H = \sum_{i,j} h_{ij} c^\dag_i c_j + \frac{1}{2}\sum_{i,j}
  U(x_i-x_j)c^\dag_{i} c^\dag_j c_j c_i,
  \label{Hamilton}
\end{equation}
where $h_{ij}$ are the matrix elements of the one-body Hamiltonian,
which depend on the chosen discretization of the second
derivative. We choose a typical spectral approximation using the
discrete Fourier transform, which also imposes periodic boundary
conditions. 
The CAP is similarly discretized as
\begin{equation}
  -i\Gamma = -i\sum_{j} \Gamma(x_j) c^\dag_j c_j, 
\end{equation}
where $\Gamma(x)$ is a non-negative function which increases
as
one approaches the boundary of the interval $[0,x_{\rm max})$ and is
zero in most of the interior.

Concerning the density operator $\rho$, we write $\rho = 
\ket{\Psi}\bra{\Psi} +
\rho_1 + \rho_0$, with $\rho_n = P_n \, \rho P_n$. Initially the system is
prepared in a two-particle pure state localized inside the absorption-free part of the grid.
The master equation for
$\rho_2(t)$ is equivalent to the usual Schr{\"o}dinger equation~(\ref{eq:schroedinger}) for $\ket{\Psi(t)}$, 
while the master equation for $\rho_1(t)$ acquires a source term, i.e.,
\begin{eqnarray}
\label{Rho1}
  i \hbar \frac{\rmd}{\rmd t}\rho_1(t) & = & [H,\rho_1(t)] - 
\{ \Gamma, \rho_1(t) \} +  \\
& &
\nonumber
2i\sum_j
  \Gamma(x_j)c_j\ket{\Psi(t)}\bra{\Psi(t)} c^\dag_j.
\end{eqnarray}

It is natural to view discrete the one-particle wave function as a vector
$\psi_1(x_j)$ of length equal to the number of discrete degrees of
freedom. Similarly, it is natural to view two-particle wave
functions as anti-symmetric matrices $\psi_2(x_j,x_k)$. Moreover, a
one-particle density matrix becomes a Hermitian matrix
$\rho_1(x,x')$. Using these notions, the master equation
\eqref{Rho1} can be compactly written as
\begin{eqnarray}
  i \hbar \frac{\rmd}{\rmd t} \rho_1(t) &=& [H, \rho_1(t)] - \{ \Gamma,
  \rho_1(t) \} +  \label{Rho1matrix} 2 i \rho_S \\
\mbox{with} \qquad \rho_S 
& \equiv & 2 h \, \psi_2^\dag \Gamma \psi_2,\nonumber
\end{eqnarray}
where the last term contains matrix-matrix products. The extra factor 2 in $\rho_S$ originates from the fact that the matrix $\psi_2$ relates to a (redundant) ``basis'' of direct product states.

If the initial two-particle state is an eigenstate of the total spin
and its projection, the source term is always proportional to a single
one-particle spin state. Hence, the spin does not introduce any
complication in the notation in this case. For parallel spins, the one
particle spin has the same direction as the two-particle spins, and
for spin projection $M_S=0$ the one-particle density operator has its
spinor component given by $(|\uparrow \rangle + (-1)^S |\downarrow
\rangle)/\sqrt{2}$, where $S$ is the total spin eigen value.

The equation for $\rho_0=p_0(t)\ket{-}\bra{-}$ becomes
\begin{equation}
  \frac{\rmd}{\rmd t}p_0(t) = \frac{2}{\hbar} \sum_j \Gamma(x_j)c_j \, \rho_1
 c_j^\dag.
  \label{eq:h0-flow}
\end{equation}
In principle, it is not necessary to include this equation in our calculations, as
$p_0(t)$ can be calculated from the constraint $\Tr[\rho(t)] = 1$.

\subsection{Collision in a Gaussian well}

\begin{figure}
\begin{center}
\epsfxsize=7.5cm \epsfbox{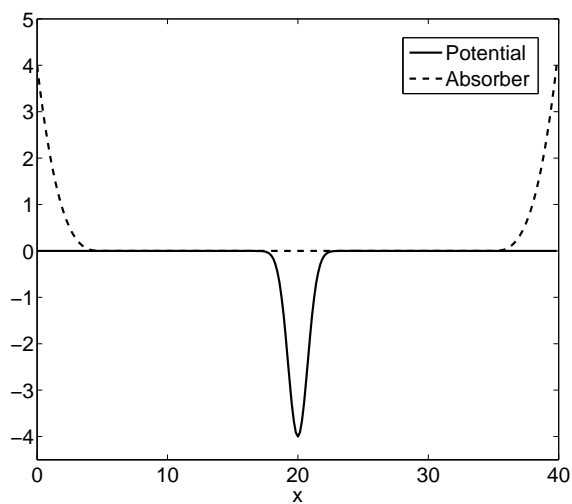}
\end{center}
\caption{The Gaussian well $V(x)$, c.f.~(\ref{potential}), and the absorber $\Gamma(x)$, c.f.~(\ref{eq:CAP}). The well has a depth of 4 and a width of 0.75, and the absorber has a third order power form for distances less than 5 from the edges at $x=0$ and $x=40$.}
\label{fig:potential}
\end{figure}

We will now focus on an example in which we set
$\hbar=m=1$ and place the electrons in a potential of Gaussian form,
\begin{equation}
\label{potential}
V(x)= -V_0 \exp \left( -\frac{(x-x_0)^2}{2\sigma^2} \right).
\end{equation}
The particles interact via a regularized Coulomb interaction
\begin{equation}
\label{interaction}
U(|x_1-x_2|)= \frac{\lambda}{\sqrt{(x_1-x_2)^2+\delta_C^2}}.
\end{equation}
For this problem we chose a CAP of power form:
\begin{eqnarray}
\label{eq:CAP}
\Gamma(x) & = & C \left(\frac{\xi}{x_T}\right)^n, \\
\nonumber
 \xi & \equiv & \max\{0,x_T-x,x-(x_{\rm max}-x_T) \},
\end{eqnarray}
where $n\geq1$ and $x_T$ is the distance from the edges at which the CAP is ``turned on'', see figure~\ref{fig:potential}.

The system may for instance serve as a model for a quantum dot which couples to the conduction band  and has narrow confinement in two dimensions.

Equation~(\ref{Rho1matrix}) is integrated using a scheme of second
order in the time step based on a standard split-step operator
scheme \cite{Feit1982}. It is instructive to consider a more general setting in order
to introduce the time stepping scheme for the density matrix. Given a
differential equation for an entity $y(t)$ in a linear space on the
form
\begin{equation}
  \dot{y}(t) = L_t y(t) + f(t),
\end{equation}
where $L_t$ is a linear operator dependent on $t$ and $f(t)$ is a
source term \emph{independent} of $y(t)$, we may integrate formally
using standard time-ordering techniques to obtain
\begin{eqnarray}
\nonumber
 y(t) & = & \mathcal{T} e^{\int_0^t L_s \rmd s} y(0)  + F(t) +
\\  & & 
\label{PropScheme}
 \sum_{n=1}^\infty \underbrace{\int_0^t \cdots
 \int_0^{s_{n-1}}}_{\text{$n$-fold}} L_{s_1} \cdots
  L_{s_n} F(s_n) \rmd s_n \cdots \rmd s_1,
\end{eqnarray}
with $F(t) = \int_{0}^t f(s)\rmd s$. The source terms are not easily
transformed using the time-ordering operator $\mathcal{T}$. Assuming
that the case $f(t)=0$ can be integrated to $p$-th order in the time
$t$ using $y(t) = \mathcal{U}_t y(0)$, a $p$-th order method for the
general case can be obtained by keeping $p$ source terms and
evaluating these with sufficiently high order quadrature. For example,
using standard Strang splitting which gives a scheme of local error
$O(t^3)$, or other schemes based on the Magnus expansion
\cite{Lubich2008}, we may approximate the term $F(t)$ as $F(t)\approx
t[f(0)+f(t)]/2$ (using trapezoidal quadrature) and the $n=1$
term as $\int_0^t L_{s} F(s) \rmd s \approx t^2 L_{t/2} [f(0) +
t \dot{f}(0)/2]$.  Specifically, in our implementation we have used a second
order scheme given by
\begin{eqnarray}
\nonumber
&&\rho_1 (t)  =  e^{-i (V+U-i\Gamma) t/2} e^{-i T t} e^{-i (V+U-i\Gamma) t/2} \rho_1(0) \times \\
\nonumber
& & 
e^{+i (V+U+i\Gamma) t/2} e^{-i T t} e^{+i (V+U+i\Gamma) t/2} +2 t \rho_S(0) +\\
& &  t^2 \left[\dot \rho_S(0) - i (H \rho_S(0) - \rho_S(0) H^\dagger) \right] + O(t^3)
\label{eq:Scheme}
\\
\nonumber
&\mbox{where}& \qquad  \dot{\rho}_S =  2 h \left(\dot{\psi}_2 \Gamma \psi_2^\dagger +  
\psi_2 \Gamma \dot{\psi}_2^\dagger \right).
\end{eqnarray}
Here $T$ is the kinetic energy.
As this scheme is trace preserving to second order only, we have also solved \eqref{eq:h0-flow} in order to check that our numerical time step is small enough to preserve the total trace.

\begin{figure}
\begin{center}
\epsfxsize=6.5cm \epsfbox{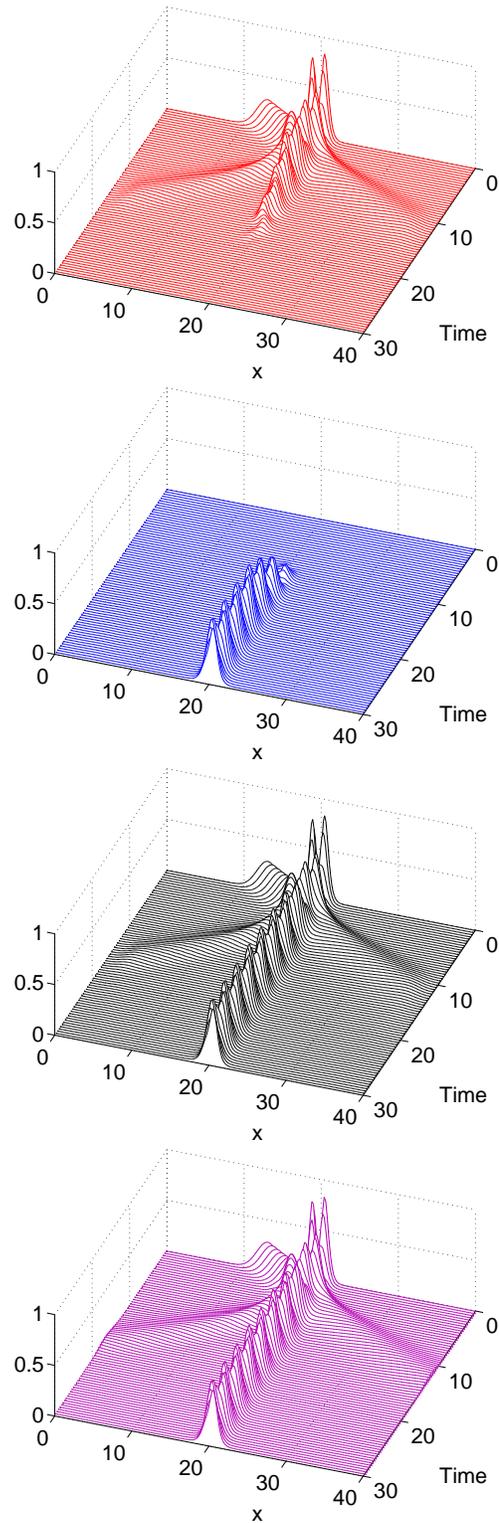}
\end{center}
\caption{Colour online: The evolution of the particle density. The upper panel is the particle density within the two-particle system, the second one from above is obtained from the one-particle density operator, and the third panel is the total particle density. Also shown, lower panel, is the particle density in the same region obtained from a solution of the full two-particle problem on a larger grid without absorber. The initial state is an anti-symmetrized state consisting of a Gaussian wave travelling towards the right with mean momentum $k_0=2$ and a state corresponding to a particle initially trapped in the well.}
\label{Evolution}
\end{figure}

Figure~\ref{Evolution} shows the evolution of the particle density for a system in which a fermion collides with another one. 
In this case the potential depth $V_0=4$ and the width $\sigma=0.75$, the interaction strength $\lambda = 5$, and the ``softening'' $\delta_C= 0.1$. The CAP has the power $n=3$, the strength $C=4$ and $x_T=5$. The starting point is a spatially anti-symmetrized state (spin triplet) consisting of a particle trapped in the well and an incoming wave packet of Gaussian shape. 
The trapped part corresponds to a superposition of the ground and the first excited one-particle states in the well, and the incoming wave function has mean momentum $k_0=2$. 
It is seen that as absorption, both due to transmission and back-scattering, takes place, the two-particle density vanishes and a one-particle density emerges. 
It is also seen that, apart from in the absorption region, the total particle density, obtained by adding the two and one-particle densities, compares well to the ``true'' particle density obtained from solving the Schr{\"o}dinger equation without absorber on a larger grid. 

\begin{figure}
\begin{center}
\epsfxsize=6.5cm \epsfbox{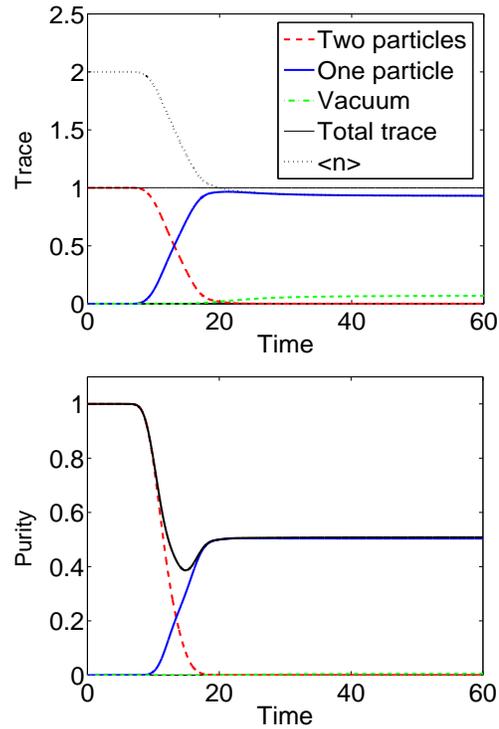}
\end{center}
\caption{Colour online: The trace and the expectation value of the number operator (upper panel) and the purity of the system (lower panel). In both cases the separate contributions from the two, one and zero part of the total density operator have been shown.}
\label{TraceEtc}
\end{figure}

Due to the collision, there is a finite probability that both particles are absorbed. 
This is clearly seen in figure~\ref{TraceEtc}, which shows how the total trace is distributed between the sub-spaces ${\cal H}_2$, ${\cal H}_1$ and ${\cal H}_0$ as a function of time. 
In this particular case we have $P(1; t \rightarrow \infty) = 0.92$ and $P(0; t \rightarrow \infty) = 0.077$.
Also shown are the expectation value of the particle number and the purity of the system. Purity is reduced in two ways. Firstly, it is reduced as the trace becomes distributed between the three sub-systems, and secondly because $\rho_1$ is not a pure state within ${\cal H}_1$. This is seen from the fact that conditional purity, $\varsigma_1$, converges towards 0.6, i.e. less than unity (not shown explicitly in the figure).

\subsection{Laser ionization of a one-dimensional helium atom}

In this next example we expose our system to an electric pulse of type
\begin{equation}
\label{Field}
E(t)=E_0 \sin^2 \left( \frac{\pi t}{T}\right) \cos(\omega t).
\end{equation}
With this time-dependent perturbation, the one-particle Hamiltonian (\ref{OnePartHam}) acquires a time-dependent term, which in the length gauge representation reads
\begin{equation}
\label{NewOnePartHam}
x E(t).
\end{equation}
We have here set the charge of the particles (the electrons) to be $-1$.
The static potential $V_N$ felt by the particles is chosen to be a regularized Coulomb potential,
\begin{equation}
\label{SmoothCoulomb}
V_N(x)=-\frac{2}{\sqrt{x^2+\delta_N^2}},
\end{equation}
whereas the interaction between them is still described by (\ref{interaction}). By choosing $\delta_N^2=\frac{1}{2}$~a.u.~and $\delta_C = 0.5735$~a.u., the ground state energy and the first ionization threshold coincide with those of a true three-dimensional atom, i.e. the ground state energy is $-2.904$~a.u.~and the ground state energy of "He$^+$" is $-2$~a.u.. By "a.u." we mean atomic units, defined by choosing the Bohr radius, the electron mass, the elementary charge and $\hbar$ as units for their respective quantities. The ground state of the system, which is a spin singlet state, is easily obtained by propagation in imaginary time.

\begin{figure}
\begin{center}
\epsfxsize=5.5cm \epsfbox{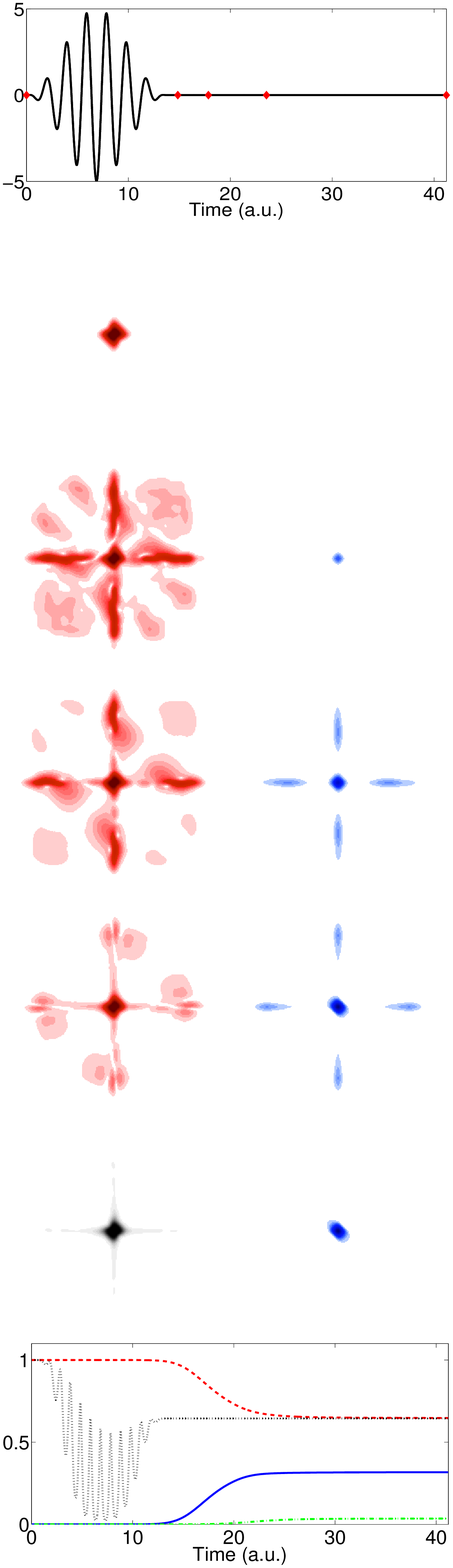}
\end{center}
\caption{Colour online: The panels show the evolution of a model one-dimensional helium atom exposed to a five cycle sine-squared electromagnetic pulse of strength $E_0=5$~a.u.~and central frequency $\omega=3.2$~a.u.~(upper panel). The middle panels show the absolute value of the wave function of the two-particle part $\psi_2(x_1,x_2)$ (left) and the absolute value of the density matrix of the one-particle part $\rho_1(x,x')$  (right) at various instances. The time of each "snapshot" is indicated by a diamond in the upper panel. The lower panel shows the probability of finding two (dashed curve), one (full curve) and zero particles (dash-dotted curve) in the system in the same manner as in the upper panel of figure~\ref{TraceEtc}. The dotted curve shows the probability of the system being in the initial state, $|\langle \psi_2(t=0)|\psi_2(t)\rangle|^2$. For all the middle panels the axes extend from $-34$ a.u.~to $+34$ a.u.~and the same colour scaling has been used in all of them.}
\label{HeliumCase}
\end{figure}

Figure~\ref{HeliumCase} shows the evolution of the system exposed to a pulse of maximum field strength $E_0=5$~a.u.,~central frequency $\omega=3.2$~a.u.~and a duration corresponding to three optical cycles. This central frequency corresponds to a photon energy which energetically allows for one photon double ionization. Rather than an absorber of power-form (\ref{eq:CAP}), we have here used a Manolopoulos-type absorber \cite{Manolopoulos2002}, which has the advantages that it is transmission free and dependent on one parameter only. Along with a figure showing how the partial traces, given by (\ref{PartialTrace}), evolve in time, and another one showing the time-dependence of the electric field, given by (\ref{Field}), we have included snapshots of the absolute values of the two-particle wave function $\psi_2(x_1,x_2)$ and the one-particle density matrix $\rho_1(x,x')$. It is clearly seen that as absorption takes place in the two-particle sub-system, a one-particle density matrix emerges. Note that the lobes following the axes of the $(x,x')$-coordinate system do {\it not} correspond to absorption of the second electron but rather loss of coherence within the one-particle sub-system.
However, from the lower panel of figure~\ref{HeliumCase}, which clearly demonstrates how single ionization may be distinguished from double ionization, we see that there {\it is} a finite probability of absorbing both particles. 
In this case, specifically, the probability of ionizing only one electron is $P(1; t \rightarrow \infty)=0.31$, and the probability of double ionization is $P(0; t \rightarrow \infty)=0.034$.

\section{Conclusion} 
\label{sec:conclusion}

In conclusion, we have demonstrated how the Lindblad equation in Fock
space can be used to generalize the notion of absorbing boundary
conditions for $N$-particle systems. Specifically, the remainder of
the system is preserved as a particle is absorbed.  
With this formalism it may be possible to describe the dynamics of
unbound systems which otherwise would require an unrealistically large
numerical grid. 

As a consequence of this being a Markovian process, some coherence is
lost, and the state after absorption is in general given by a mixed
state rather than a wave function.  
Within Lindblad theory, the identification between the Lindblad operators and the creation and annihilation operators comes quite natural for complex absorbing potentials. We believe that the method outlined in this paper may be generalized to other kinds of non-Hermitian ``interactions'' as well.

Since it is considerably more involved to solve a master equation
rather than a Schr{\"o}dinger equation, c.f.~\cite{Molmer1992},
future work will aim to find lower rank methods for solving \eqref{eq:block-flow}. Also, more sophisticated spatial approximations like sparse grids \cite{Lubich2008} may be utilized to simplify the treatment of more particles.

\ack
Constructive suggestions from Michael Genkin is gratefully acknowledged by the authors.

\begin{thebibliography}{10}

\bibitem{Taylor1996}
Jonathan Parker, K~T Taylor, Charles~W Clark, and Sayoko Blodgett-Ford.
\newblock Intense-field multiphoton ionization of a two-electron atom.
\newblock {\em J. Phys. B}, 29:L33, 1996.

\bibitem{Krausz2007}
M~Uiberacker, Th. Uphues, M.~Schultze, A.~J. Verhoef, V.~Yakovlev, M.~F. Kling,
  J.~Rauschenberger, N.~M. Kabachnik, H.~Schröder, M.~Lezius, K.~L. Kompa,
  H.-G. Muller, M.~J.~J. Vrakking, S~Hendel, U.~Kleineberg, U.~Heinzmann,
  M.~Drescher, and F.~Krausz.
\newblock Attosecond real-time observation of electron tunnelling in atoms.
\newblock {\em Nature}, 446:627, 2007.

\bibitem{Kosloff1986}
R.~Kosloff and D.~Kosloff.
\newblock Absorbing boundaries for wave-propagation problems.
\newblock {\em J. Comput. Phys.}, 63(2):363--376, 1986.

\bibitem{Fevens1999}
Thomas Fevens and Hong Jiang.
\newblock Absorbing boundary conditions for the schr\"{o}dinger equation.
\newblock {\em SIAM J. Sci. Comput.}, 21(1):255--282, 1999.

\bibitem{Antoine2008}
Xavier Antoine, Anton Arnold, Christophe Besse, Matthias Ehrhardt, and Achim
  Schaedle.
\newblock A review of transparent and artificial boundary conditions techniques
  for linear and nonlinear schrodinger equations.
\newblock {\em Comm. Comput. Phys.}, 4(4):729--796, 2008.

\bibitem{Berenger1994}
JP~Berenger.
\newblock {A perfectly matched layer for the absorption of
  electromagnetic-waves}.
\newblock {\em {J. Comput. Phys.}}, {114}({2}):{185--200}, {OCT} {1994}.

\bibitem{Manolopoulos2002}
David~E. Manolopoulos.
\newblock Derivation and reflection properties of a transmission-free absorbing
  potential.
\newblock {\em J. Chem. Phys.}, 117(21):9552--9559, 2002.

\bibitem{Monnerville1999}
M~Monnerville and JM~Robbe.
\newblock Optical potential discrete variable representation method in the
  adiabatic representation - application to the co(b-1 sigma(+)-d `(1)sigma(+),
  j=0) predissociation process.
\newblock {\em Eur. Phys. J. D}, 5(3):381--387, 1999.

\bibitem{Grozdanov2006}
TP~Grozdanov, L~Andric, and R~McCarroll.
\newblock Calculations of partial cross sections for photofragmentation
  processes using complex absorbing potentials.
\newblock {\em J. Chem. Phys.}, 124(9), 2006.

\bibitem{Karlsson2009}
Hans~O. Karlsson.
\newblock Stability of the complex symmetric lanczos algorithm for computing
  photodissociation cross sections using smooth exterior scaling or absorbing
  potentials.
\newblock {\em J Phys. B}, 42:125205, 2009.

\bibitem{Kulander1987}
Kenneth~C. Kulander.
\newblock Multiphoton ionization of hydrogen: A time-dependent theory.
\newblock {\em Phys. Rev. A}, 35(1):445--447, 1987.

\bibitem{Protopapas1996}
M~Protopapas, CH~Keitel, and PL~Knight.
\newblock Relativistic mass shift effects in adiabatic intense laser field
  stabilization of atoms.
\newblock {\em J. Phys. B}, 29(16):L591--L598, 1996.

\bibitem{Lindblad1976}
G.~Lindblad.
\newblock On the generators of quantum dynamical semigroups.
\newblock {\em Comm. Math. Phys.}, 48:119, 1976.

\bibitem{Schlosshauer}
M.~Schlosshauer.
\newblock {\em Decoherence and the quantum-to-classical transition}.
\newblock Springer-Verlag, 2007.

\bibitem{Sandulescu1987}
A~Sandulescu and H~Scutaru.
\newblock Open quantum-systems and the damping of collective modes in deep
  inelastic-collisions.
\newblock {\em Ann. Phys.}, 173(2):277--317, 1987.

\bibitem{Caban2005}
Pawe\l{} Caban, Jakub Rembieli\ifmmode~\acute{n}\else \'{n}\fi{}ski, Kordian~A.
  Smoli\ifmmode~\acute{n}\else \'{n}\fi{}ski, and Zbigniew Walczak.
\newblock Unstable particles as open quantum systems.
\newblock {\em Phys. Rev. A}, 72(3):032106, 2005.

\bibitem{Bertlmann2006}
Reinhod~A. Bertlmann, Walter Grimus, and Beatrix~C. Hiesmayr.
\newblock Unstable particles as open quantum systems.
\newblock {\em Phys. Rev. A}, 73(5):054101, 2006.

\bibitem{Feit1982}
M.D Feit, J.A~Fleck Jr., and A~Steiger.
\newblock Solution of the schrödinger equation by a spectral method.
\newblock {\em J.Comput. Phys.}, 47(3):412 -- 433, 1982.

\bibitem{Lubich2008}
C.~Lubich.
\newblock {\em From Quantum to Classical Molecular Dynamics: Reduced Models and
  Numerical Analysis}.
\newblock European Mathematical Society, 2008.

\bibitem{Molmer1992}
Jean Dalibard, Yvan Castin, and Klaus M\o{}lmer.
\newblock Wave-function approach to dissipative processes in quantum optics.
\newblock {\em Phys. Rev. Lett.}, 68(5):580--583, 1992.

\end{thebibliography}

\end{document}